# Confinement suppresses the effect of rotation on convection

Vinay Kumar Tripathi[1], Pranav Joshi[1*]

[1]*Fluid and Thermal Systems Laboratory, Department of Mechanical Engineering, Indian Institute of Technology Kanpur, Kanpur 208016, India*

**ABSTRACT**. We perform direct numerical simulations (DNS) to study the effect of bi-lateral confinement, i.e., the effect of aspect ratio, on the Rayleigh number (Ra) for the onset of wall-mode convection ($Ra_{wm}$) in rotating Rayleigh-Bénard convection (RBC) for various Ekman number E ($10^{-2} \leq E \leq 10^{-5}$) and aspect ratio $\Gamma$ ($0.08 \leq \Gamma \leq 5$). For a given E, as the aspect ratio is lowered from a large value, $Ra_{wm}$ initially decreases slowly, reaching a minimum at $\Gamma = \Gamma_{min}$ ($\Gamma_{min} \sim E^{0.09}$). As $\Gamma$ is decreased further below $\Gamma_{min}$, $Ra_{wm}$ increases rapidly and for sufficiently strong confinement, i.e., $\Gamma < \Gamma_*$ ($\Gamma_* \sim E^{1/3}$), becomes indistinguishable from $Ra_c$, the critical Rayleigh number for non-rotating RBC, at the same $\Gamma$. We designate the ranges $\Gamma_{min} < \Gamma < 5$ as the 'confinement-affected' regime and $\Gamma < \Gamma_*$ as the 'spatially constrained' regime. For a given rotation rate, the spatially constrained regime extends to higher Ra as the aspect ratio is decreased. We propose that in this regime, strong lateral confinement suppresses the horizontal velocity, and hence the Coriolis force, rendering rotation ineffective over most of the domain.

## I. INTRODUCTION.

Natural convection observed in various geophysical and astrophysical flows, e.g., the Earth's atmosphere and outer core [1,2], inside the ocean, and Jupiter's atmosphere, has been widely modeled using a canonical system, the Rayleigh-Bénard convection (RBC), in which fluid is heated from below and cooled from the top under background rotation with axis anti-parallel to gravity. Rotating RBC (RRBC) is characterized by the Rayleigh number (Ra = $\alpha g \Delta T H^3 / \nu \kappa$), which denotes the strength of the buoyancy forcing, the Prandtl number (Pr = $\nu/\kappa$), which is a fluid property, the Ekman number (E = $\nu/2\Omega H^2$), which indicates the strength of the rotation rate (a higher rotation rate means a lower value of E), and the aspect ratio ($\Gamma$) which denotes the ratio of the horizontal dimension ($L$) to the vertical dimension ($H$) of the RBC cell. Where $\Delta T$, $g$, $\rho$, $\kappa$, $\nu$, $\alpha$ are the temperature difference between the bottom and top walls, the acceleration due to gravity, the density, the thermal diffusivity, the kinematic viscosity, and the isobaric thermal expansion coefficient of the fluid, respectively. The Nusselt number (Nu =< $1 + \sqrt{RaPr}[\iiint U_z T. dV]/V$ >) denotes the heat transfer efficiency. Here, $U_z$, $V$, and $T$ are the dimensionless vertical velocity, the total volume of the RBC cell, and the dimensionless temperature, respectively and <> denotes time avergaing.

In most geophysical phenomena of interest, convection occurs under strong buoyancy forcing and substantial effects of rotation. To reach closer to the geophysical and astrophysical regimes of rotating convection, i.e., large Ra and low E, experimental studies must incorporate large $H$ and slender RBC cells, i.e., low $\Gamma$. Thus, such experimental studies e.g., [3–6] are likely to suffer from the effects of lateral confinement by no-slip sidewalls, which are evident from the differences shown by the heat transfer and flow morphology as compared with those for laterally unbounded domains [7].

Thus, the present study focuses on the effect of bidirectional lateral confinement with no-slip insulating sidewalls on rotating RBC to delineate the effects of the aspect ratio on the heat transfer and the flow structure. Specifically, we study the low-Ra regime of RRBC and focus on the effect of the aspect ratio on the onset of wall mode convection ($Ra_{wm}$). We show that at sufficiently small aspect ratios, i.e., $\Gamma < \Gamma_*$, in the 'spatially constrained regime' (to be discussed in what follows) of rotating convection, the strong lateral confinement renders rotation ineffective so that the onset of convection under rotation occurs at the same Ra as that without rotation.

For non-rotating RBC, the onset of convection in a laterally infinite domain with no-slip horizontal boundaries occurs at a critical Rayleigh number of $Ra_c^\infty \approx 1708$ [8]. The superscript $\infty$ indicates the aspect ratio of infinity for a laterally unbounded domain. The variation of $Ra_c^\Gamma$ with $\Gamma$ ($Ra_c^\Gamma$ indicates $Ra_c$ as a function of $\Gamma$) for laterally confined cells with all no-slip boundaries has been reported by several prior studies, e.g., [9–12]. Note that the degree of side wall confinement in rectangular cuboid domains can be increased either by decreasing one of $\Gamma_x$ and $\Gamma_y$, fixing the other (termed unidirectional or one-dimensional (1D) confinement) [13,14], or by decreasing both $\Gamma_x$ and $\Gamma_y$ equally, i.e., $\Gamma_x = \Gamma_y = \Gamma$ (termed bidirectional or two-dimensional (2D) confinement), with only the latter possible ($\Gamma_r = \Gamma$) for a circular cylindrical domain [15]. Here, $\Gamma_x = L_x/H$, $\Gamma_y = L_y/H$, and $\Gamma_r = D/H$, where $L_x$ and $L_y$ are the horizontal extents of the rectangular cuboid along the $x$ and $y$ directions, respectively, and $D$ is the diameter of the circular cylinder.

---

*Contact author: jpranavr@iitk.ac.in, tvinay@iitk.ac.in

Using a variational approach, Shishkina [11] reported $Ra_c^\Gamma \approx (2\pi)^4(1 + \Gamma_x^{-2})(1 + \Gamma_x^{-2}/4 + \Gamma_y^{-2}/4)$ for rectangular cuboids having no-slip insulating sidewalls with $\Gamma_y \leq \Gamma_x \equiv \Gamma$, while $Ra_c^\Gamma \approx (2\pi)^4(1 + 1.49\Gamma^{-2})(1 + 0.34\Gamma^{-2})$ for a circular cylinder. Ahlers et al. [15], using linear stability analysis (LSA) and direct numerical simulation (DNS), showed that for cylindrical containers $Ra_c^\Gamma = Ra_c^\infty(1 + C\Gamma^{-2})^2$ with $C \lesssim 1.49$. The recent experimental study in a cylindrical container by Ren et al. [12] reported $Ra_c^\Gamma = 915\Gamma^{-4.05}$ for $\Gamma \leq 1/10$. Although, in general, $Ra_c^\Gamma$ increases with an increase in both 1D and 2D lateral confinement, Wagner and Shishkina [14] report for 1D confinement a non-monotonic dependence of $Ra_c^\Gamma$ on $\Gamma$, e.g., $Ra_c^{1/10} \approx 10^6$ and $Ra_c^{1/4} \approx 3 \times 10^6$.

Rotation suppresses the motion and delays the onset of convection in RRBC, i.e., increases $Ra$ for the onset. For a laterally unbounded domain, prior studies report $Ra_c = (8.7 - 9.6E^{1/6})E^{-4/3}$ [16–18] for no-slip horizontal boundaries. However, in the presence of no-slip adiabatic side walls, the onset of convection is advanced relative to that for a laterally unbounded domain and occurs at $Ra = Ra_{wm}$, $Ra_{wm} < Ra_c$ [18–21]. Advanced convection starts near the side walls in alternate hot and cold structures, which are widely known as wall modes [18,20,22]. Prior asymptotic studies report $Ra_{wm}^\infty = \pi^2\sqrt{6\sqrt{3}}E^{-1} + 46.55E^{-2/3}$ [20] for no-slip horizontal boundaries, the superscript $\infty$ indicating large $\Gamma$. The value of $Ra_{wm}^\infty$ is obtained from the leading order asymptotic approximation and is valid for low Ekman number and sufficiently large aspect ratios [23].

However, only one prior study, by Goldstein et al. [24], shows that the Rayleigh number for the onset decreases with a decrease in aspect ratio from a large value $\Gamma = 4$, and increases as the aspect ratio is decreased beyond a sufficiently small $\Gamma$ for *shear-free top and bottom and no-slip sidewalls*. Thus far, no study known to the author reports $Ra_{wm}$ as a function of $\Gamma$ for all no-slip walls.

## II. METHODS

Direct numerical simulations are performed in a rotating cylinder of a square cross-section ($\Gamma \times \Gamma \times 1$). The governing equations (Eq. 1-3), non-dimensionalized using the free fall velocity $U = \sqrt{g\alpha\Delta TH}$, $H$, and $\Delta T$, are solved using a GPU-accelerated finite difference solver pySaras [25], which is discussed in the supplementary material of Anas and Joshi [25] and at a greater depth in Mohammad Anas's dissertation [26]. No-slip boundary condition is used for all walls, while isothermal and adiabatic conditions are used for horizontal and side walls, respectively. We have performed all the simulations at $Pr = 0.7$ (corresponding to most gases) and for Ekman number $E = \infty, 10^{-2}, 6 \times 10^{-4}, 10^{-4}, 10^{-5}$. For all the simulations in this work, the number of grid points in the vertical (horizontal) direction lies between 64 (32) and 512 (512). A minimum of 5 grid points lie within the thinnest boundary layer, e.g., the Ekman boundary layer on the horizontal walls for low Ra, which is deemed to provide sufficient resolution for present purposes [27].

$$\boldsymbol{\nabla} \cdot \boldsymbol{u} = 0. \quad (1)$$

$$\frac{\partial \boldsymbol{u}}{\partial t} + (\boldsymbol{u} \cdot \boldsymbol{\nabla})\boldsymbol{u} = -\nabla p + \left(\sqrt{\frac{Pr}{Ra}}\right)\nabla^2 \boldsymbol{u} + T\hat{\boldsymbol{e}}_z - \left(\sqrt{Pr/RaE^2}\right)\hat{\boldsymbol{e}}_z \times \boldsymbol{u} \quad (2)$$

$$\partial T/\partial t + (\boldsymbol{u} \cdot \boldsymbol{\nabla})T = (1/\sqrt{RaPr})\nabla^2 T \quad (3)$$

Here, $\boldsymbol{u}, p, T$, and $\hat{\boldsymbol{e}}_z$ are the non-dimensional velocity, non-dimensional pressure, non-dimensional temperature, and a unit vector along the rotation axis, respectively.

## III. RESULTS

In Fig.1, we show the variation of Nu with Ra for various combinations of E and $\Gamma$. For non-rotating RBC, Ra for the onset, i.e., $Ra_c^\Gamma$, is a function of $\Gamma$, which increases as $\Gamma$ is decreased from a large $\Gamma \approx 5$. Immediately beyond the onset, the rate of increase in Nu with Ra, i.e., the exponent $\beta$ in Nu$\sim$Ra$^\beta$, increases as the confinement becomes stronger. For a given $\Gamma$, $\beta$ decreases with an increase in Ra and eventually becomes equal to that for large $\Gamma = 5$ beyond a sufficiently large Ra, i.e., classical scaling $Nu = 0.15Ra^{0.29}$ [28,29].

Similarly to confinement, rotation also delays the onset of convection: for a fixed aspect ratio, as the Ekman number decreases, the onset of convection is delayed [8,18]. Note that Ra for the onset of motion with no-slip sidewalls, even for a very large aspect ratio, i.e., $Ra_{wm}^\infty$, is lower than that for laterally unbounded domains, $Ra_c$. This is because the wall mode convection begins near the no-slip sidewalls at a Rayleigh number lower than $Ra_c$, keeping the bulk devoid of convection. With the increase in Ra, convection also occurs in the central region, i.e., away from the sidewalls.

*Contact author: jpranavr@iitk.ac.in, tvinay@iitk.ac.in

For given $\Gamma$ and E, $\beta$ is highest near the onset of convection. As Ra is increased further in the rotationally constrained (RC) regime [3,18,30,31], $\beta$ decreases and becomes equal to that for non-rotating RBC at that $\Gamma$ for sufficiently large Ra in the rotation-unaffected regime. For a given E, $Ra_{wm}$ is a function of $\Gamma$: as the aspect ratio is decreased from a large value, $Ra_{wm}^{\Gamma}$ initially decreases slightly before increasing substantially, approaching $Ra_c^{\Gamma}$ (without rotation), as $\Gamma$ is decreased below a small value that depends on E. This aspect will be discussed further shortly. For a given E, $\beta$ in the RC regime also varies with $\Gamma$. Under a sufficiently strong bi-lateral confinement, Nu at low Ra becomes equal to that for non-rotating RBC at that $\Gamma$. This convergence between the values of Nu for rotating and non-rotating RBC persists up to a Rayleigh number that seems to be a function of both E and $\Gamma$.

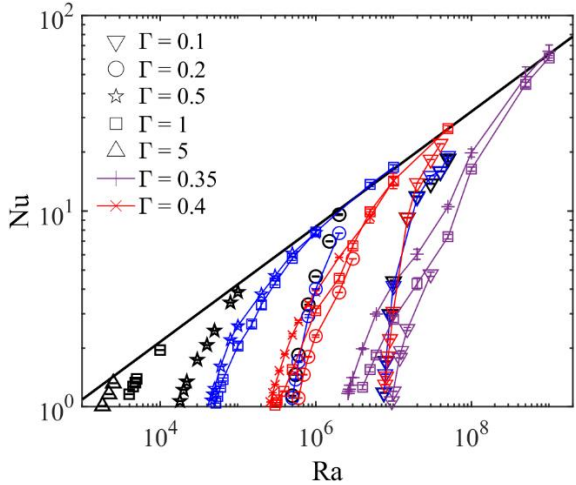

FIG 1. Variation of Nu with Ra for various E and $\Gamma$ at Pr = 0.7. Colors denote the Ekman number: black, E = $\infty$; blue, E = $6 \times 10^{-4}$; red, E = $10^{-4}$; and purple, E = $10^{-5}$. Here, the black solid line denotes Nu = $0.15 Ra^{0.29}$ for present data at $\Gamma = 5$ and E = $\infty$, for $10^4 \leq Ra \leq 5 \times 10^7$. Note that to retain clarity, the data for $\Gamma > 1$ is not shown for RRBC.

To quantify the effect of $\Gamma$ on the Rayleigh number for the onset of wall mode convection, i.e., $Ra_{wm}^{\Gamma}$, we estimate $Ra_{wm}^{\Gamma}$ by fitting a power law between Nu and Ra for $\epsilon^{\Gamma} = Ra/Ra_{wm}^{\Gamma} \lesssim 1.3$ and extrapolating the same to Nu = 1 for each $\Gamma$ at a given E. In Fig. 2(a), we show that for a given E, $Ra_{wm}^{\Gamma}$ slightly decreases with a decrease in $\Gamma$ from $\Gamma = 5$. We refer to the value of $\Gamma$ at which $Ra_{wm}^{\Gamma}$ reaches a minimum as $\Gamma_{min}$. Upon further increasing the confinement, i.e., as $\Gamma$ decreases below $\Gamma_{min}$, $Ra_{wm}^{\Gamma}$ increases sharply and, for sufficiently small aspect ratios, becomes indistinguishable from $Ra_c^{\Gamma}$, the critical Rayleigh number for non-rotating convection. We refer to the aspect ratio below which $Ra_{wm}^{\Gamma} \approx Ra_c^{\Gamma}$ as $\Gamma_*$. Note that $\Gamma_*$ decreases with a decrease in E, i.e., as the strength of rotation is increased, a stronger confinement is required for the Rayleigh number for convection onset to be unaffected by rotation. Goldstein et al. [24] observed for shear-free top and bottom walls an initial decrease followed by an increase in $Ra_{wm}$ as aspect ratio is decreased. However, the present results for the case of *no-slip* top and bottom walls further show that, importantly, $Ra_{wm}$ becomes the same as $Ra_c$ below a certain $\Gamma(E)$.

For E = $6 \times 10^{-4}$ (for which we have a sufficient range of aspect ratios lower than $\Gamma_*$), the present data follows the power law $Ra_{wm}^{\Gamma} = 1235.7 \Gamma^{-3.75 \pm 0.2}$ for $\Gamma \leq \Gamma_*$ ($\Gamma_* \approx 0.35$). For non-rotating RBC, we observe a similar exponent in the power law, $Ra_c^{\Gamma} = 1148.2 \Gamma^{-3.78 \pm 0.14}$, over the aspect ratio range $0.1 \leq \Gamma \leq 0.5$. This result for non-rotating RBC is in good agreement with that of Ahlers et al. [15], who reported $Ra_c^{\Gamma} = 1202 \Gamma^{-3.74 \pm 0.02}$ in the range $0.1 \leq \Gamma \leq 0.35$ for non-rotating RBC (see Fig. 2(a)). The data for other Ekman numbers, being indistinguishable from $Ra_c^{\Gamma}$ below their respective $\Gamma_*$, see Fig. 2(a), also seem to suggest a similar power law.

To emphasize the dependence of $Ra_{wm}$ on $\Gamma$, and hence, clearly observe the minimum in $Ra_{wm}$, the inset in Fig. 2(a) shows the variation with the aspect ratio of the Rayleigh numbers for the onset of convection, $Ra_{wm}^{\Gamma}$ and $Ra_c^{\Gamma}$, normalized by their respective values at $\Gamma = 5$. It seems that, similarly to $\Gamma_*$, $\Gamma_{min}$ also decreases with decreasing E, albeit at a much slower rate. However, although $\Gamma_*$ seems to exist for all E, the same is not true for $\Gamma_{min}$: the data for the largest Ekman number, E = $10^{-2}$, do not show a clear minimum in $\widetilde{Ra}_{wm}$.

In Fig. 2(b), we show the variation with E of $Ra_{wm}^{\Gamma}/Ra_c^{\Gamma}$, i.e., a measure of the impact of rotation on the onset of convection in finite cells having no-slip sidewalls. As expected, for all aspect ratios, $Ra_{wm}^{\Gamma}/Ra_c^{\Gamma}$ increases as the Ekman number decreases. However, $Ra_{wm}^{\Gamma}/Ra_c^{\Gamma}$ decreases consistently with a decrease in $\Gamma$ indicating that the growing effect of spatial confinement counters that of rotation on the onset of convection. $Ra_{wm}^{\Gamma}/Ra_c^{\Gamma}$ is expected to start increasing beyond 1 only when the Ekman number is decreased below a certain value, say $E_0$ (see, e.g., [32]). The present results show that $E_0$ decreases as $\Gamma$ is reduced, i.e., as confinement is increased, stronger rotation is required to delay the onset of convection beyond that for the corresponding non-rotating case.

*Contact author: jpranavr@iitk.ac.in, tvinay@iitk.ac.in

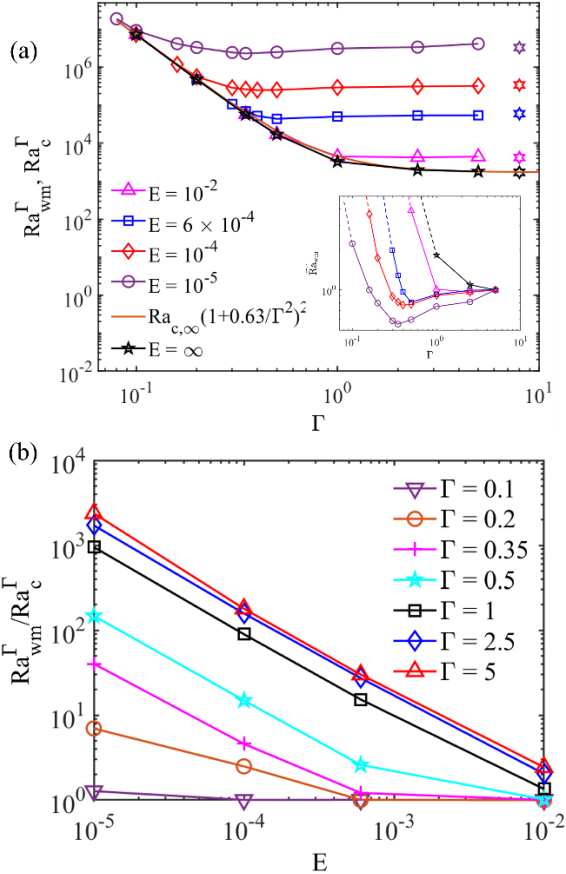

FIG 2. Variation of (a) $Ra_c^\Gamma$ and $Ra_{wm}^\Gamma$ with $\Gamma$ at various E and (b) $Ra_{wm}^\Gamma/Ra_c^\Gamma$ with E at various $\Gamma$. The inset in (a) shows the variation of $\widetilde{Ra}_{wm} = Ra_{wm}^\Gamma/Ra_{wm}^5$ with $\Gamma$. Here, hexagonal stars denote $Ra_{wm}^\infty$ and $Ra_c^\infty$ at that respective Ekman numbers from Zhang et al. [20]. Note that the expression $Ra_c^\infty(1 + 0.63/\Gamma^2)^2$ [15] is considered to retain the continuity of $Ra_c^\Gamma$ with $\Gamma$ for non-rotating RBC. Note that for $0.1 \leq \Gamma \leq 0.35$, $Ra_c^\Gamma = 1202\Gamma^{-3.74\pm0.02}$ [15].

## IV. DISCUSSION

The relative effects of rotation and confinement on the Rayleigh number for the onset of convection can be summarized in a regime diagram shown in Fig. 3, which presents the variation of $\Gamma_{min}$ and $\Gamma_*$ with the Ekman number. We designate $\Gamma \approx 5$ as the rough lower boundary of the confinement-unaffected (CuA) regime of rotating convection since the degree of confinement is not expected to have any significant effect on $Ra_{wm}^\Gamma$ at large aspect ratios. The regime between $\Gamma \approx 5$ and $\Gamma_{min}$ is the confinement-affected (CA) regime of rotating convection, in which $Ra_{wm}^\Gamma$ decreases as the aspect ratio is decreased and reaches a minimum at the regime boundary $\Gamma = \Gamma_{min}$ ($\Gamma_{min} \sim E^{0.09}$). In this regime, the effects of rotation are still strong enough to result in $Ra_{wm}^\Gamma$ clearly higher than $Ra_c^\Gamma$. As $\Gamma$ is decreased further below $\Gamma_{min}$, $Ra_{wm}^\Gamma$ starts increasing rapidly as the flow transitions to the spatially constrained (SC) regime with $\Gamma_* \sim E^{0.33}$ as its upper boundary. In the SC regime, $Ra_{wm}^\Gamma$ becomes indistinguishable from $Ra_c^\Gamma$, i.e., the effects of rotation on convection onset are not evident. As the Ekman number increases, the range of aspect ratios for the transition from the CA regime to the SC regime decreases, with $\Gamma_{min}$ approximately the same as $\Gamma_*$ at sufficiently low rotation rates, $E > \sim 2 \times 10^{-3}$. Note that the observed trend $\Gamma_* \approx 4.42E^{0.33}$ makes a reasonable estimation that the SC regime grows up to $\Gamma \approx 5$ at $E \sim O(1)$, i.e., when a negligible effect of rotation is expected at any aspect ratio. Figure 3 also shows $l_c/H \approx 2.4E^{1/3}$, the normalized horizontal length scale at the onset for $Pr \gtrsim 0.68$ [7,17]. Interestingly, the spatially constrained regime occurs for $\Gamma < \Gamma_* \approx 2(l_c/H)$.

*Contact author: jpranavr@iitk.ac.in, tvinay@iitk.ac.in

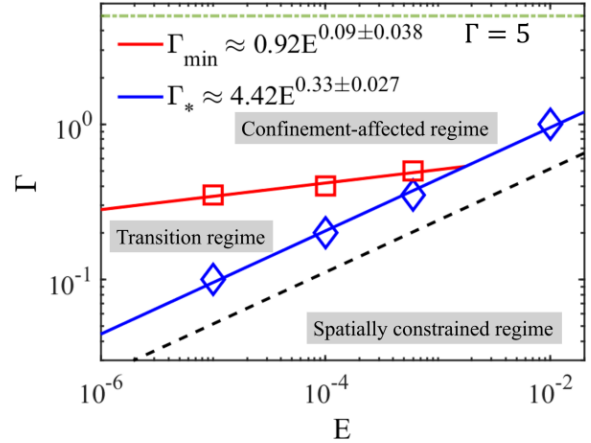

FIG 3. Map of various confinement regimes in the $E - \Gamma$ space. Here, the black dashed line denotes $l_c/H = 2.4E^{1/3}$ [8,18]. The red squares and blue diamonds denote $\Gamma_{min}$ and $\Gamma_*$, respectively, while the corresponding solid lines denote the power-law fit to the data.

As $\Gamma$ is decreased below $\Gamma_*$, not only are the Rayleigh numbers at onset equal for rotating and non-rotating convection, but the Nusselt numbers also become equal up to greater values of $\epsilon^\Gamma$ (see Fig. 1, at $E = 6 \times 10^{-4}$ for $\Gamma_c = 0.2$ and $\Gamma = 0.1$). Thus, at a given E, the spatially constrained regime extends to larger Rayleigh numbers as the aspect ratio is decreased. To demonstrate the effect of confinement on the velocity field in this regime, Fig. 6 shows the variation of $U_r = <U_h>/<U_z>$ with $\Gamma$ for various E, where $U_h$ and $U_z$ are the horizontal and vertical velocities, respectively, and $<>$ denotes the volumetric average. We observe that when $\Gamma > \Gamma_{min}$, the effect of confinement on $U_r$ is not significant and $U_r \sim O(1)$, suggesting isotropy at the global level. However, as the confinement increases, i.e., $\Gamma$ decreases below $\Gamma_{min}$, the horizontal velocity is suppressed to a greater extent than the vertical one. Consequently, $U_r$ decreases following a power law $U_r = \alpha(E)\Gamma^\psi$ (refer to the caption of Fig. 4), where the pre-factor $\alpha$ is a function of E. While $\psi \approx 1.5$ for all E, $\alpha \sim E^{-1/2}$, suggesting that a stronger confinement is required for the same global anisotropy ($U_r$) as rotation becomes stronger (E decreases). This observation also agrees with $\Gamma_*$, the upper boundary of the SC regime, decreasing with decreasing E. As the horizontal velocity is suppressed with increasing confinement, we can expect the Coriolis force, and hence the effects of rotation, to diminish as well since the Coriolis force is dependent on the horizontal velocity and not the vertical.

velocity, and hence the Coriolis force, is expected to be greater away from the horizontal walls. For the present data, $1/E_L \sim O(0.1)$ in the bulk when $\Gamma \sim \Gamma_*$, indicating the weakening of the Coriolis force relative to the viscous force as the system enters the SC regime. Thus, as $\Gamma$ is decreased significantly, rotation is rendered ineffective over most of the domain as the horizontal velocity, and hence the Coriolis force, is suppressed by the lateral confinement. Thus, with negligible effect of the Coriolis force in the spatially constrained regime ($\Gamma \lesssim \Gamma_*$), $Nu(E) \approx Nu(E = \infty)$ is observed up to a certain $\epsilon^\Gamma$. With an increase in $\epsilon^\Gamma$, the buoyancy force increases, which also contributes to the Coriolis force [33], thus $Nu(E)$ deviates from $Nu(E = \infty)$ beyond a certain $\epsilon^\Gamma$. For the convergence between $Nu(E)$ and $Nu(E = \infty)$ at larger $\epsilon^\Gamma$, confinement must be further increased to counter the rotation effect.

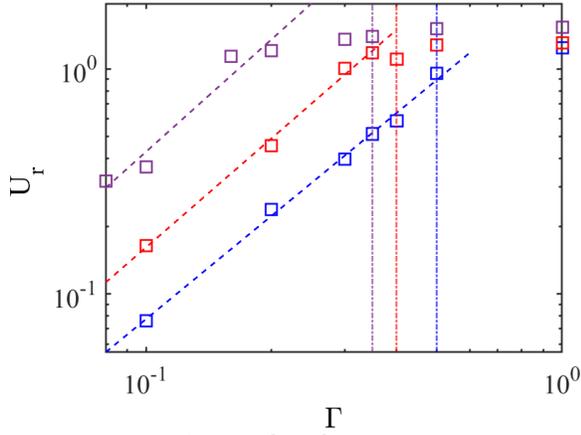

Fig.4 Variation of $U_r$ with $\Gamma$ for various E. Here, blue dashed line: $U_r = \alpha(E)\Gamma^{1.52 \pm 0.1}$; red dashed line: $U_r = \alpha(E)\Gamma^{1.61 \pm 0.15}$; purple dashed line: $U_r = \alpha(E)\Gamma^{1.66 \pm 0.55}$, and $\alpha = 0.067 E^{-0.5 \pm 0.025}$. Here, the vertical dash-dotted lines denote $\Gamma = \Gamma_{min}$ corresponding to that color, i.e., E.

To infer the relative importance of rotation close to the onset of convection, we estimate the local Ekman number ($E_L = F_V/F_c$) for a fixed $Ra/Ra^\Gamma_{wm} \approx 1.2$ and various E and $\Gamma$. Here, viscous force $F_V = (\sqrt{Pr/Ra})|\nabla^2 \boldsymbol{u}|$, and Coriolis force $F_c = (\sqrt{Pr/RaE^2})|\hat{e}_z \times \boldsymbol{u}|$. Figure 5 shows the variation of $1/E_L$ at mid-height and close to the top/bottom walls with $\Gamma$ for various E and a fixed $Ra/Ra^\Gamma_{wm} = 1.2$. As the aspect ratio decreases, $1/E_L$ decreases everywhere in the domain, following a power law when $\Gamma < \Gamma_{min}$. Note that for a given E, $\Gamma$ and $Ra/Ra^\Gamma_{wm}$, $1/E_L$ is higher close to the horizontal walls than at mid-height (or the 'bulk') since the suppression of the horizontal fits, $\Gamma = \Gamma_{\min}$, and $\Gamma = \Gamma_*$, respectively. Note that the exponents in the power law are shown in the figures.

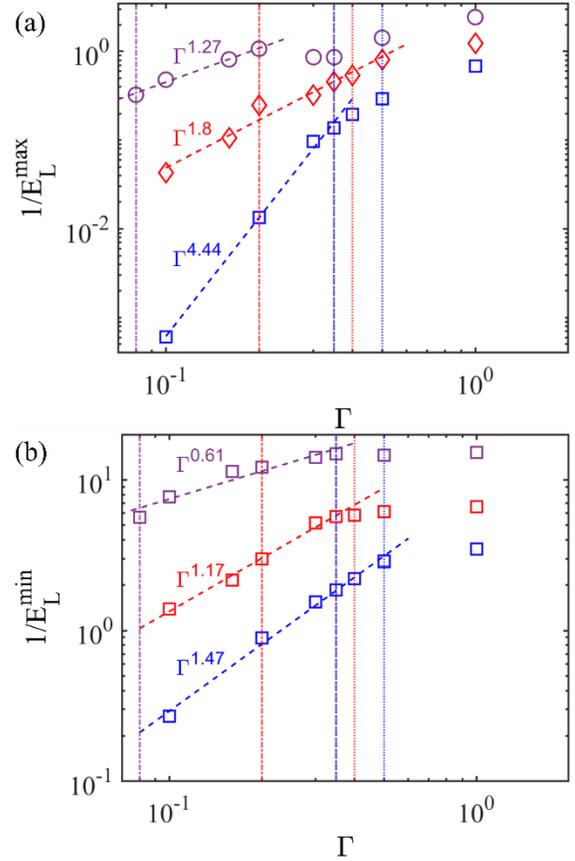

Fig.5 Variation with the aspect ratio of the inverse of (a) the local Ekman number at the mid-plane and (b) the local Ekman number near the horizontal walls for various $E$. Here, blue: $E = 6 \times 10^{-4}$, red: $E = 10^{-4}$, purple: $E = 10^{-5}$. The dashed, dotted, and dash-dotted lines denote the power law

*Contact author: jpranavr@iitk.ac.in, tvinay@iitk.ac.in

To get insights into the effect of 2D confinement on the flow structure in RRBC, we show in Fig. 6 snapshots of temperature and vertical vorticity ($\Omega_z$) for various $\Gamma$ at $E = 6 \times 10^{-4}$ and $\epsilon^\Gamma \approx 1.2$. The classical wall mode structures, similar to those observed in numerous previous studies [18,22,34], are clearly observed at large $\Gamma = 2.5$ in the form of pairs of hot and cold regions in tandem adjoining the sidewalls, accompanied by two regions of cyclonic vorticity for each pair. As the aspect ratio decreases, the number of wall mode pairs decreases until $\Gamma_{\min}$, at which a single pair of hot and cold regions accompanied by a pair of cyclonic regions is observed. Further decrease in the aspect ratio below $\Gamma_*$ results in the merging of the regions of cyclonic vorticity accompanied by a rapid decrease in the vorticity magnitude.

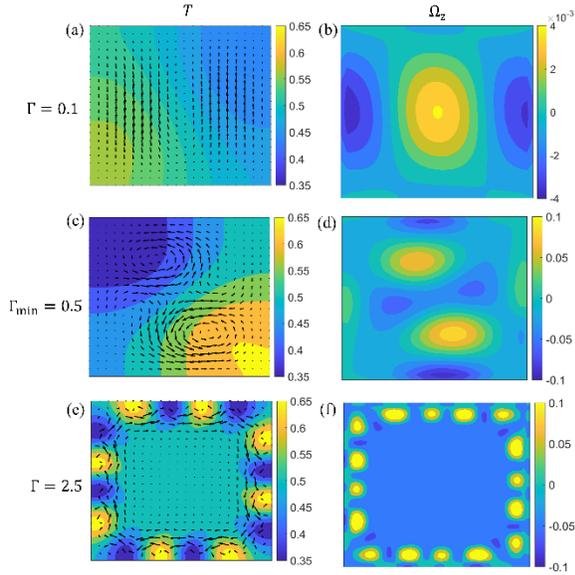

Fig.6 Temperature (first column) and vertical vorticity (second column) contours in mid-plane at $E = 6 \times 10^{-4}$ and $\epsilon^\Gamma \approx 1.2$ for various aspect ratios. Here, vectors in the first column denote the horizontal velocity and their length denotes the velocity magnitude.

Figures 7 and 8 compare the flow structures for rotating convection ($E = 6 \times 10^{-4}$) in the SC regime at $\Gamma = 0.1 < \Gamma_*$ with those of non-rotating convection at the same $\epsilon^\Gamma \approx 1.2$, i.e., when $Ra_{wm}$ and Nu are the same for the two cases. The flow structures in the vertical mid-plane (see Fig. 7), especially the vertical velocity distributions primarily contributing to the heat transfer, are indistinguishable for the two flows. The structure in the horizontal mid-plane (Fig. 9)
*Contact author: jpranavr@iitk.ac.in, tvinay@iitk.ac.in

shows some differences, however, in both cases, the horizontal velocity (and the vertical vorticity) is negligible in comparison to the vertical velocity (and horizontal vorticity). Thus, although Coriolis force (associated with the horizontal velocity) exists in the rotating system, it is too weak to affect the heat transfer.

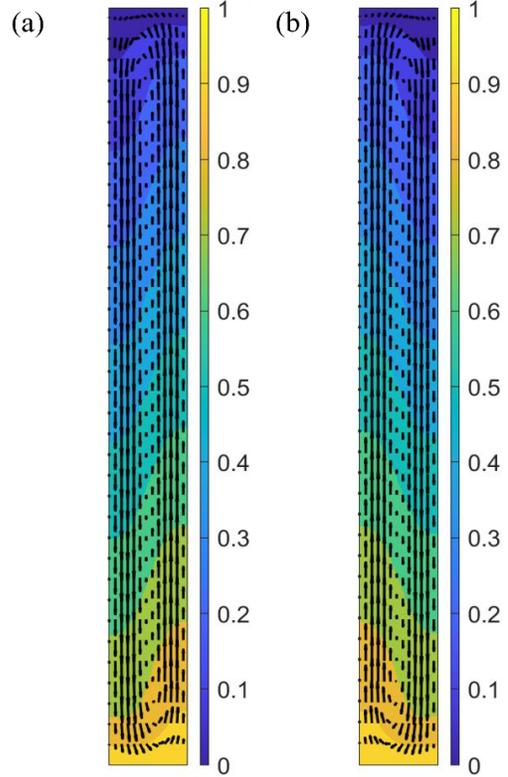

Fig.7 Temperature contours for (a) $E = \infty$ and (b) $E = 6 \times 10^{-4}$, at $\epsilon^\Gamma \approx 1.2$ and $\Gamma = 0.1$. The vectors denote the in-plane velocity and their length the velocity magnitude.

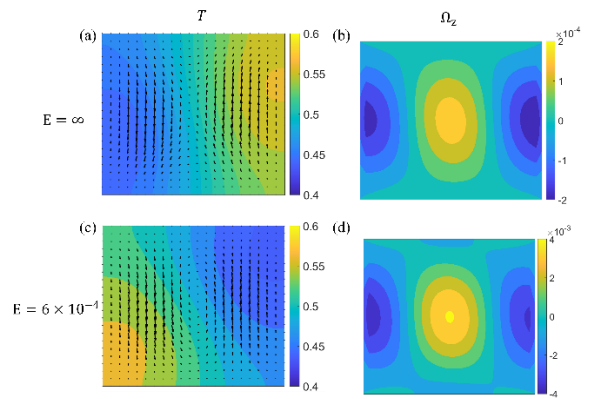

Fig.8 Temperature (a, c) and vertical vorticity (b, d) contours in mid-plane at $\epsilon^\Gamma \approx 1.2$ and $\Gamma = 0.1$ for $E = \infty$ and $E = 6 \times 10^{-4}$. Note that the color bar suggests the maximum and minimum values of $\Omega_z$ for $E = \infty$ is about 1/20 times of that for $E = 6 \times 10^{-4}$.

## V. Conclusion

The present study shows that bilateral confinement profoundly affects rotating convection. Specifically, under sufficiently strong confinement, i.e. in the spatially constrained regime, the Nusselt number for rotating convection is indistinguishable from that for non-rotating convection. Starting from the onset of convection, this cessation of the effects of rotation on the heat transfer spreads to higher Rayleigh numbers as the aspect ratio is made smaller. We propose that strong lateral confinement suppresses the horizontal velocities that are solely responsible, along with the background rotation, for the presence of the Coriolis force. The present results are potentially important for future studies using smaller aspect ratios to reach geophysically relevant parameter ranges and set the platform for further exploring the regime boundaries at lower Ekman numbers and higher Rayleigh numbers.

## ACKNOWLEDGMENTS

The numerical simulations are performed on PARAM Sanganak, the supercomputing facility at IIT Kanpur. We also thank Prof. M.K. Verma and his group for developing the finite difference code SARAS. We thank Mohammad Anas for developing the GPU-accelerated finite difference-based Python solver. V.K.T. thanks Mohammad Anas for helping him in the initial learning of the new GPU-accelerated solver. This work was funded by the Science and Engineering Research Board (SERB), Department of Science & Technology, Government of India, project no. SRG/2019/001037 and CRG/2022/008483.

*Contact author: jpranavr@iitk.ac.in, tvinay@iitk.ac.in

*Contact author: jpranavr@iitk.ac.in, tvinay@iitk.ac.in